\documentclass[preprint,12pt]{elsarticle}

\usepackage{amssymb}
\usepackage{graphicx}  
\usepackage{subfigure} 
\usepackage{wrapfig}  
\newcommand{\cci}{$CeCoIn_5$}

\journal{Physica C}

\begin{document}

\begin{frontmatter}



\title{Josephson effect in a {\cci} microbridge}


\author{O Foyevtsov, F Porrati and M Huth}

\address{Physikalisches Institut, Goethe Universit\"at Max-von-Laue-Strasse 1 60438 
Frankfurt am Main, Germany}

\begin{abstract}
We report DC Josephson effects observed in a microbridge prepared from an individual crystalline growth domain of {\cci} thin film. Josephson effects were observed by periodic voltage modulations under external magnetic field $\Delta V(B)$ with the expected periodicity and by the temperature dependence of the Josephson critical current $I_c(T)$. The shape of $\Delta V(B)$ was found to be asymmetric, as it is expected for microbridges. The dependence $I_c(T)$ follows the Ambegaokar-Baratoff relation, which is unexpected for microbridges. Features in the dynamical resistance curves were attributed to the periodic motion of Abricosov vortices within the microbridge. 
\end{abstract}


\end{frontmatter}


\section{Introduction}
Strong correlation effects give rise to the intermediate valence phenomena in the heavy fermion (HF) compounds. These intrasite correlations are governed by the compactness of the $f$-shells of rare earth ions, the central component of most of the HF materials. It is these properties which are responsible for the rich variety of ground states found in the HF compounds, many of which are $Ce$-based structures. One particularly interesting example is the superconducting (SC) ground state discovered in $CeCu_2Si_2$ \cite{steglich1979superconductivity} at ambient pressure with $T_c=0.6~K$. Many of its isostructural compounds were also found to become SC under pressure demonstrating the proximity to a quantum critical point. Another exciting class of $Ce$-based compounds is the so-called $115$ family, e.g. $CeMIn_5$ compounds, where $M$ denotes one of the transition metals $Co$, $Rh$ or $Ir$. Our investigations focus on the SC properties of {\cci}. Superconductivity in {\cci} emerges at a rather high temperature as for a heavy fermion of $T_c=2.3~K$ from a non-Fermi liquid state \cite{nakajima2007non}. Additionally, the theoretically predicted Fulde\mbox{-}Ferrel\mbox{-}Larkin\mbox{-}Ovchinnikov state was observed in {\cci} in sufficiently large magnetic fields \cite{bianchi2003possible}. The SC state in {\cci} is unconventional. Measurements of the electronic specific heat in {\cci} suggest the $d_{xy}$ symmetry of the SC order parameter (OP), while point contact spectroscopy and thermal conductivity studies favor the $d_{x^2-y^2}$ symmetry of the OP \cite{park2009andreev}. This discrepancy may be solved by planar tunneling studies, which were previously successfully applied in thin films of $UPd_2Al_3$ and $UNi_2Al_3$ (Reference~\cite{PlanarUPdAl,zakharov2006tunneling,Jourdan1997335}). However, only a limited number of HF materials may be obtained in the form of epitaxial thin films, a prerequisite for creating reliable planar tunneling junctions. As it was shown by several groups, thin films of {\cci} and $CeIn_3$ suffer from a discontinuous surface morphology making preparation of planar tunneling experiments exceedingly difficult \cite{0953-8984-19-5-056006,Zaitsev200952,izaki:122507,Foyevtsov20107064}. Nevertheless, recent studies showed that micron-sized {\cci} thin film crystalline growth domains may become accessible to electronic transport measurements by using modern microstructuring techniques \cite{Foyevtsov20107064}. Further improvements of these techniques allowed us to investigate the SC properties of {\cci} on a SQUID device prepared from an individual growth domain \cite{PhysRevB.84.045103}. One has to point out that the Josephson effects were successfully applied for the determination of the OP in cuprates \cite{PhysRevLett.71.2134} and were also applied to bulk single crystals of {\cci} and $CeIrIn_5$ \cite{sumiyama2003josephson,fukui2003surface}.

In this work we report the second part of our investigations of the SC properties of {\cci} by Josephson effects. In the first paper we reported on the {\cci} microcrystalline SQUID preparation and characterization \cite{PhysRevB.84.045103}, here we focus on investigations of a single {\cci} microbridge demonstrating Josephson effects.
\section{Experimental}
\begin{figure}[tb] 
\center
\includegraphics[width=13cm]{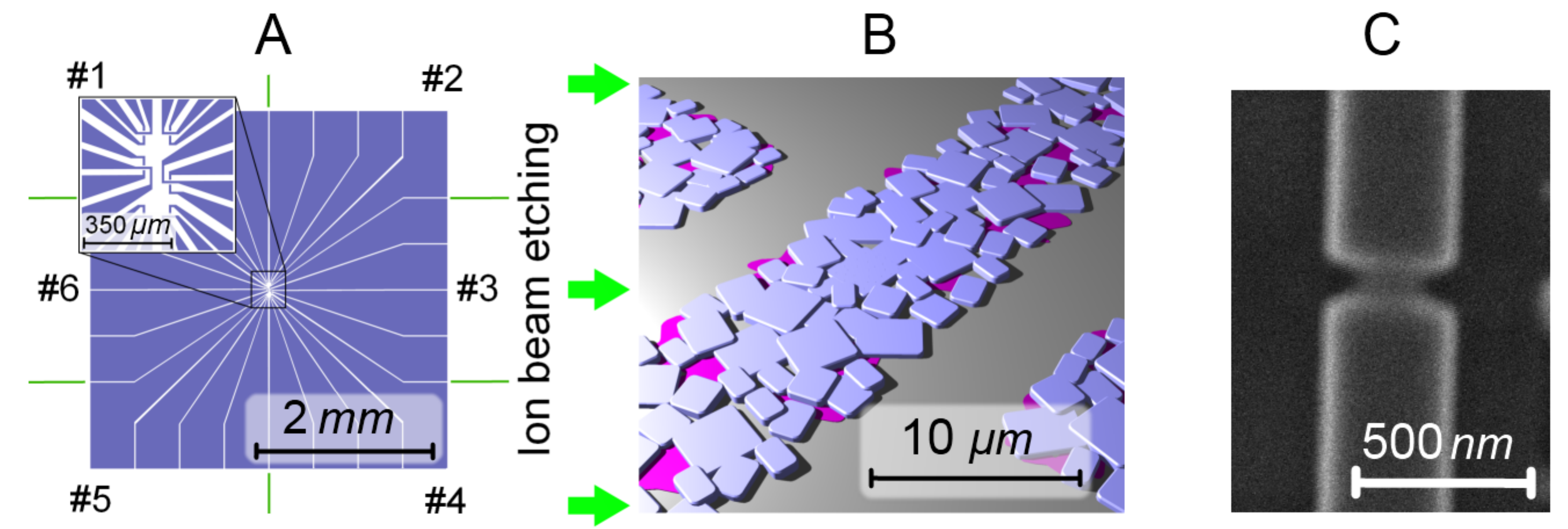} 
\caption{(A) Schematic view of the photo mask used in a positive photolithographic process for partitioning the {\cci} thin film into a number of electrically isolated current-carrying paths, which form six identical patterns for further microstructuring. (B) Schematic view of the {\cci} interconnected crystallite paths of one of the six patterns. An appropriate {\cci} crystallite from the central path was then patterned via focused ion beam milling in an SEM. (C) An SEM micrograph of the microbridge. The microbridge length $L=90~nm$ is the distance separating the two SC banks, the width $W=250~nm$, the height $h\approx 150~nm$, and the hight of the {\cci} CS banks $d=300~nm$.}
\label{Mask}  
\end{figure}

{\cci} thin films were grown on $10\times 10~mm^2$ $Al_2O_3~(11\bar{2}0)$ substrates by molecular beam epitaxy method, as it was described in more details earlier \cite{0953-8984-19-5-056006,Foyevtsov20107064}. The resulting films' morphology is discontinuous and consists of partially overlapping $c$-axis oriented {\cci} growth domains or microcrystals of few microns average lateral dimensions. The thickness of the films, as measured by atomic force microscopy, is approximately $300~nm$. For addressing a single {\cci} microcrystallite a thin film was pre-patterned with ultraviolet photolithography and ion beam etching defining six patterns for further microstructuring, as is schematically shown in Figure~\ref{Mask}. Initially, one of the {\cci} crystallites was used for structuring a SQUID with two microbridges, as it was reported in \cite{PhysRevB.84.045103}. The microstructuring was performed by using focused ion beam etching and focused ion beam induced deposition from $W(CO)_6$ precursor gas in an FEI Nova NanoLab 600 dual beam scanning electron microscope (SEM) \cite{huth2009conductance,10.1116/1.3196789}.
The transport characteristics of the {\cci} microbridges during preparation were monitored \textit{in situ} in the SEM chamber, as described in details in \cite{0957-4484-20-19-195301}. The low temperature electronic transport measurements were performed in a top-loading $^3He$ cryostat. All electrical measurements were performed with a four probe technique either using a voltage or current driven transport measurement setup or using a standard setup for dynamic resistance measurements with a Stanford SR830 Lock-In amplifier.

The single microbridge structure (see Figure~\ref{Mask}(C)) was obtained by permanently suppressing the Josephson coupling in one of the microbridges via increasing current density through the initial SQUID loop. As a result, due to a small asymmetry between the cross-sections of the two SQUID microbridges the Josephson coupling through the weaker bridge was permanently suppressed.
\section{Results and discussions}
\begin{figure}[tb] 
\center
\includegraphics[width=8cm]{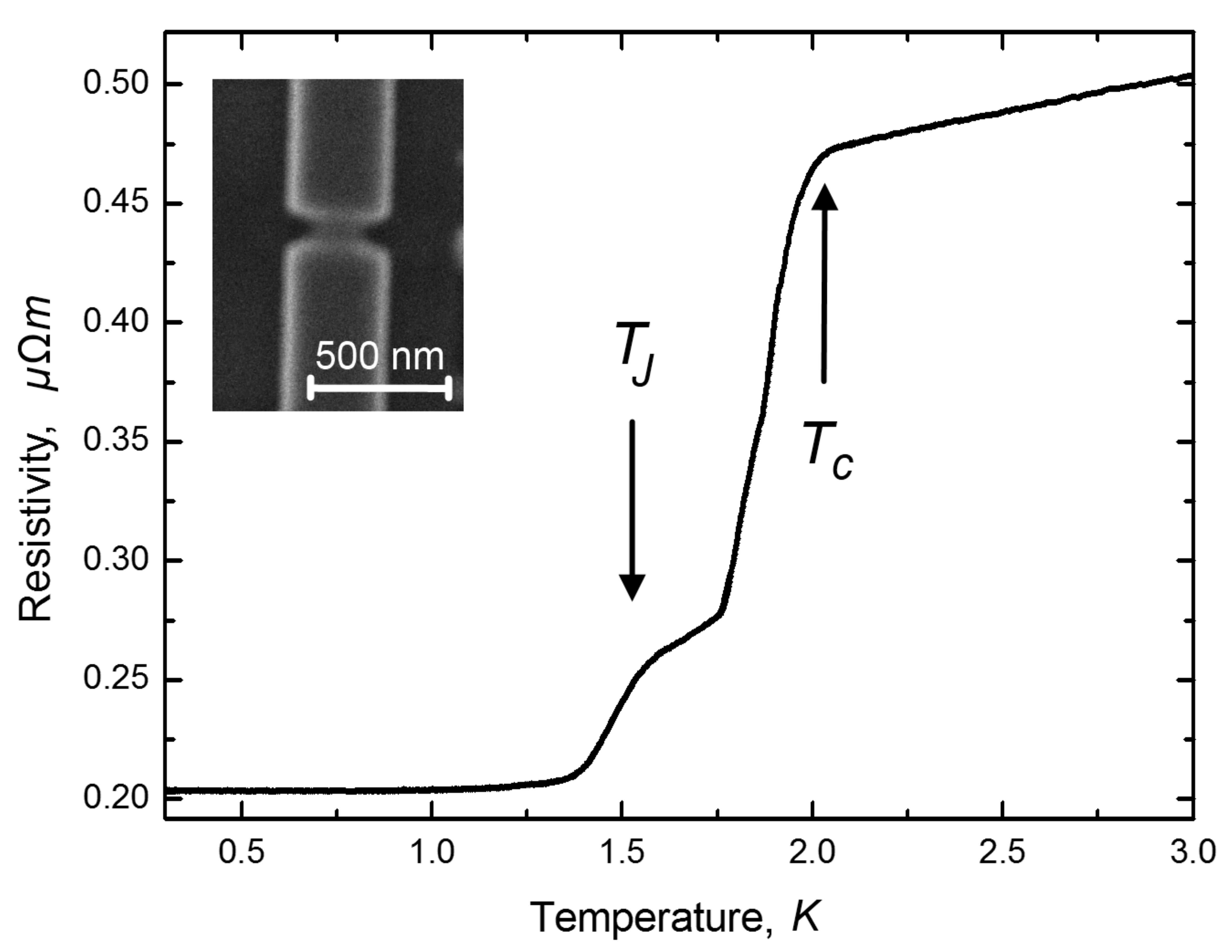} 
\caption{A low temperature part of the temperature dependence of resistivity $\rho(T)$ measured during cooling down with a constant current of $0.5~\mu A$. The arrows denote the superconducting transition temperature $T_c$ of {\cci} and the Josephson coupling temperature $T_J$.}
\label{RvsT}  
\end{figure}

The temperature dependence of the resistivity $\rho(T)$ of the microbridge was measured in the range $0.3-300~K$ during the cool down. In the heavy fermion state of {\cci} $\rho(T)$ demonstrats its typical behaviour with a residual resistance ratio of $RRR_{300K/2.5K}\approx 1.73$. This value is lower than that obtained for the SQUID structure, which we attribute to the rather intrusive transformation of the SQUID into the single bridge structure that involved high current densities. The low temperature part of $\rho(T)$ for the microbridge is shown in Figure~\ref{RvsT}. The superconducting transition of {\cci} was found at $T_c\approx 2~K$, which is a typical value reported for thin films and microcrystals of this compound \cite{0953-8984-19-5-056006,Zaitsev200952,Haenisch2010S568}.
The additional transition at about $1.5~K$ is associated with the Josephson coupling temperature $T_J$ \cite{PhysRevB.84.045103}. 

The resistivity of the microbridge structure at $2.5~K$ is approximately $0.47~\mu\Omega m$, which is deduced when a single branch in the equivalent electrical scheme of the initial SQUID is considered. This value is about $9\%$ larger than that for the SQUID structure studied in \cite{PhysRevB.84.045103}, which is consistent with qualitative expectations. Moreover, when this increase of resistivity is assigned solely to the constriction area, the weakest part of the structure, the resistivity must increase by about $300\%$, if no asymmetry between the initial SQUID bridges is assumed. Another remarkable feature is the relatively large residual resistivity $\rho_0=0.2~\mu\Omega m$ of the microbridge below $T_J$. While a much smaller value of $\rho_0$ of the SQUID was previously associated with the Ambegaokar and Halpherin mechanism \cite{PhysRevLett.22.1364}, here, this mechanism cannot be justified. In this case the large value of $\rho_0$ can be regarded as the increased weight of the dissipative transport channels within the microbridge. It should be stressed that in spite of the reduced $RRR$ and suggested substantial increase of the resistivity within the microbridge, no noticable change of $T_J$ with regard to our previous analysis \cite{PhysRevB.84.045103} is observed. We may speculate that its value is strongly influenced by the anomalously large electronic mean free path in {\cci} at low temperatures \cite{PhysRevB.72.214515, movshovich2001unconventional, ikeda2001unconventional}, which still could be large when compared to the microbridge length.

\begin{figure}[tb] 
\center
\includegraphics[width=9cm]{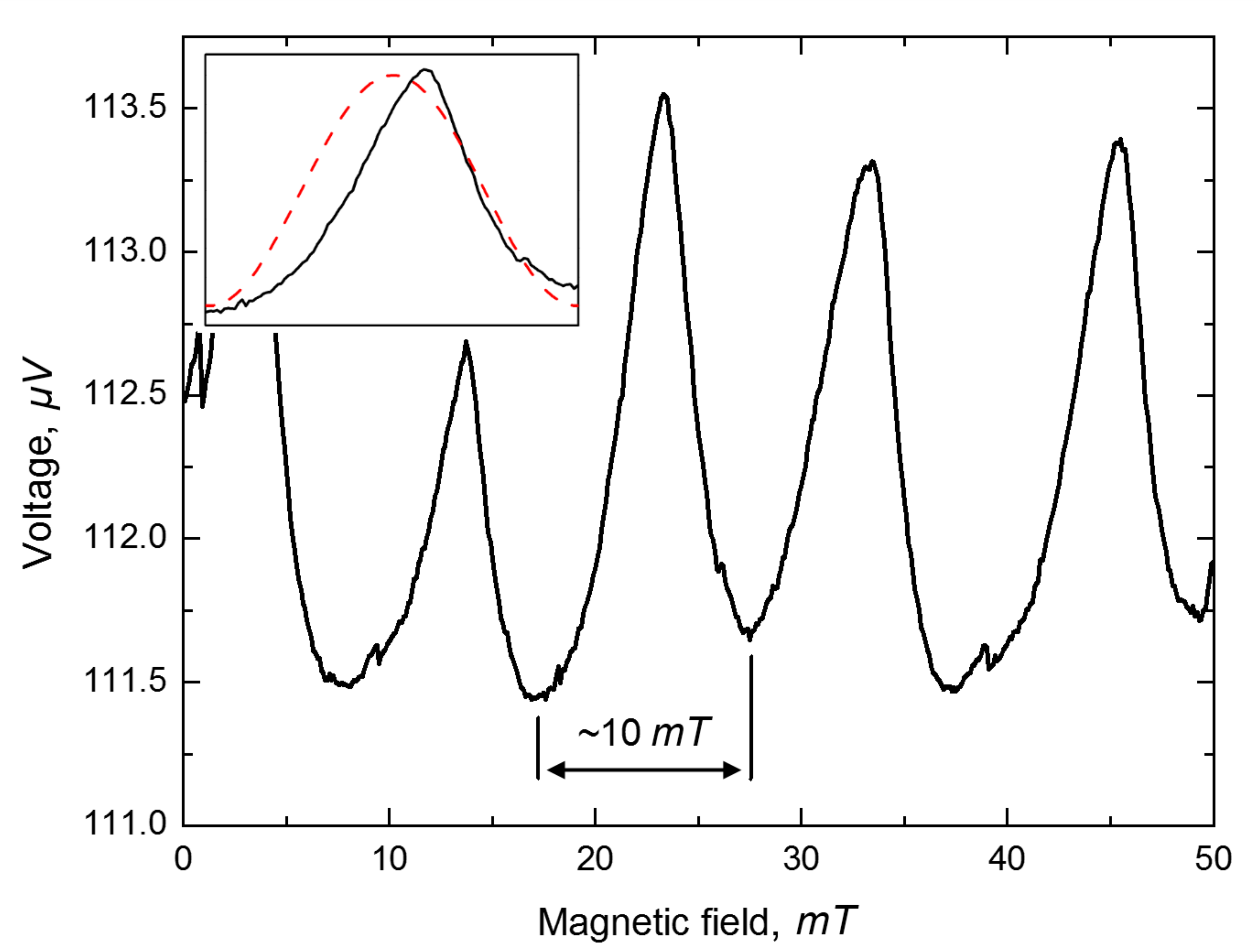} 
\caption{Typical microbridge voltage modulations by the external magnetic field measured at $T=0.3~K$ with constant current $I=8~\mu A>I_c$. The inset shows a comparison between the shapes of the measured modulations (solid line) and theoretically expected sinusoidal modulations for an ideal tunneling Josephson junction (dashed line).}
\label{Vmod}  
\end{figure}
Josephson effects are usually recognized by the Shapiro steps in the $V\mbox{-}I$ characteristics under microwave radiation or by the modulation of the Josephson critical current $I_c$, or, similarly, by voltage modulations $\Delta V$ under variable external magnetic field $B$ applied perpendicularly to the junction/microbridge plane. The periodicity of these $\Delta V(B)$ modulations is given by $\Delta B=\Phi_0/s$, where $\Phi_0$ is the flux quantum and $s$ is the effective area of the microbridge. For microbridges demonstrating a Josephson effect the effective area is given as $s=[W(2\lambda+L)]$, where $W$ and $L$ are the microbridge width and length, and $\lambda$ is the Josephson penetration depth \cite{rosenthal1991flux}. The effective area $s$ is small for microbridges when compared to the SQUID and, hence, $\Delta V(B)$ has a larger period $\Delta B$. In the main part of Figure~\ref{Vmod} we present the measured $\Delta V(B)$ modulations for the single {\cci} microbridge, which we ascribe to the Josephson voltage modulations. In our previous work the observed long period modulations of the SQUID were ascribed to the individual Josephson microbridges, although these modulations were substantially anharmonic \cite{PhysRevB.84.045103}. We related this anharmonicity to the interference effects between the two Josephson microbridges connected via the {\cci} SC banks. 

The harmonicity of the $\Delta V(B)$ modulations is much improved in the single microbridge supporting our assumption of the strong influence of interference effects in the SQUID structure. The period length of the modulations is $10~mT$, which is in good agreement with the estimated microbridge effective area if $\lambda=235~nm$ is used \cite{0295-5075-62-3-412}. In contradistinction to the previous work, no small period superimposed SQUID voltage modulations were detected, which also confirms the single microbridge geometry. Another feature is the pronounced asymmetry of the $\Delta V(B)$ modulations shown in the inset of Figure~\ref{Vmod}, which was not observed before for the SQUID. Such deviations from the ideal Josephson effect in microbridges were reported before for various types of weak links \cite{RevModPhys.76.411,RevModPhys.51.101}. There it was shown that the ratio between the electronic mean free path $l$, the Ginsburg-Landau superconducting coherence length $\xi$, and the microbridge dimensions $L,W$ defines different transport regimes and that in some of these regimes the voltage modulations can be asymmetric. According to the predictions, when those regimes are established within a microbridge, another essential characteristic, the temperature dependence of the Josephson critical current $I_c(T)$, must also deviate from its ideal behaviour in tunneling junctions. However, one should mention that the specific properties of {\cci}, such as the very large $l$ and small $\xi$ \cite{PhysRevLett.97.127001} as well as the microbridge dimensions $L,W$ do not allow for a strict classification of our microbridges in terms of the regimes mentioned above.	

\begin{figure}[tb] 
\center
\includegraphics[width=8cm]{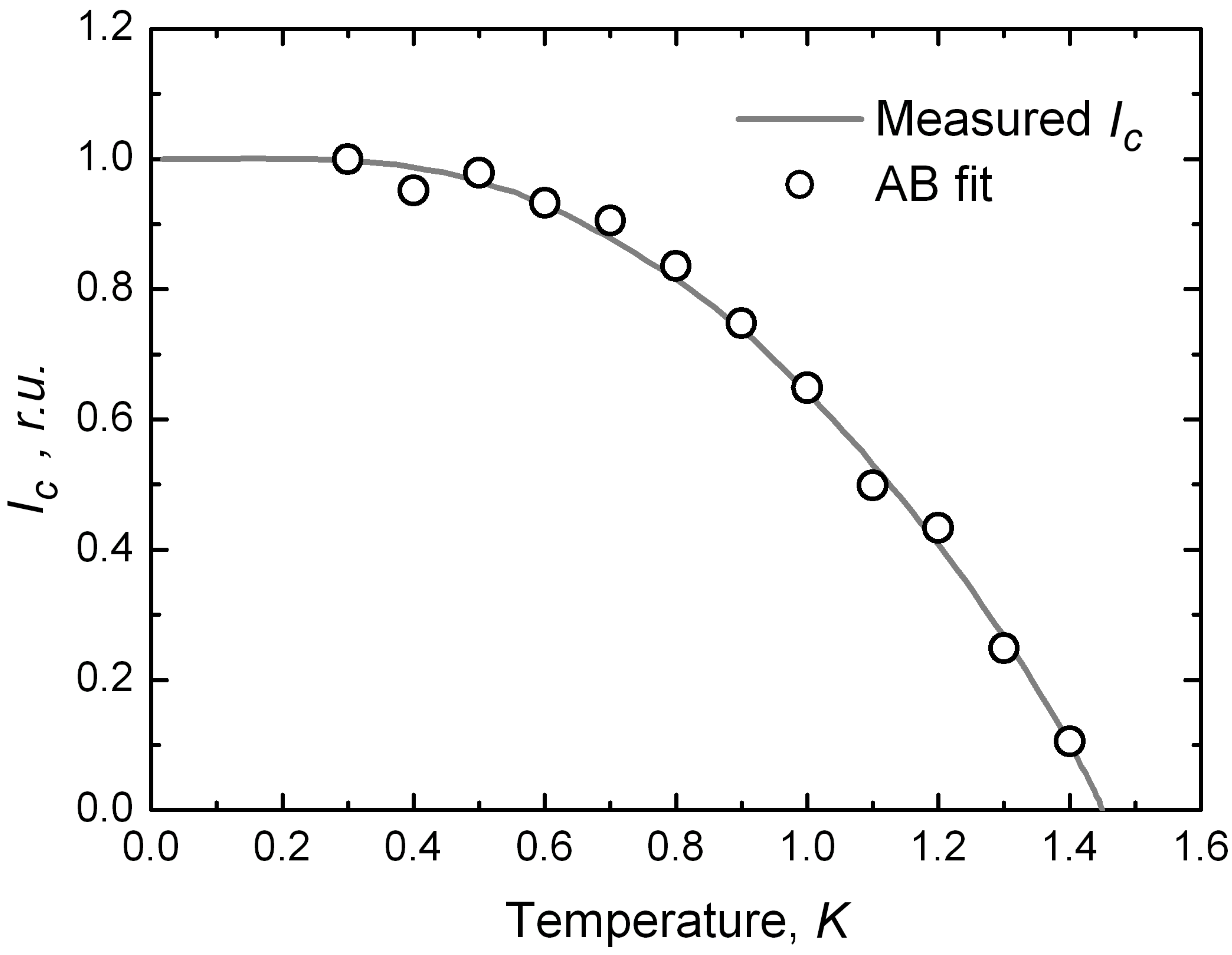} 
\caption{Josephson critical current measured as a function of temperature (circles) and the Ambegaokar-Baratoff fit (solid line) with $R_n^{fit}=71~\Omega$ and $T_c=1.45~K$.}
\label{ICvsT}  
\end{figure}

The measured temperature dependence of the Josephson critical current $I_c(T)$ is shown in Figure~\ref{ICvsT} by open circles. The solid line is the Ambegaokar-Baratoff (AB) fit to the measured data. The function $I_c(T)$ was reported for the {\cci} SQUID in our previous work, where it also followed the AB model. 
In both cases, the expected behaviour of $I_c(T)$ derived for microbridges, like in the Kulik-Omelyanchuk models for the diffusive or ballistic regimes \cite{KO.1975,KO.1977} or in the Ishii's model \cite{PTP.44.1525} for long bridges, is not observed. In contrast, the AB model is derived in the tunneling limit with small barrier transparency and without coordinate dependencies of the SC parameters within the barrier. This suggests that the requirement $l\gg L$ is fulfilled in this case. In the AB fit of $I_c(T)$ for the SQUID the values of the local $T_c$ and $T_J$ were not discriminated. For the single microbridge we find a local $T_c=1.45~K$ according to the best AB fit, which is lower than $T_J=1.55~K$ derived from the $\rho(T)$ curve. On one hand, a reduced value of the local $T_c$ with respect to the SQUID results should not be surprising due to the modified transport properties of the microbridge during preparation. On the other hand, the emerging difference between the values of $T_J$ and the local $T_c$ for the single microbridge suggests different processes associated to them. The value of the normal state resistance $R_n^{fit}$ in the AB fit is found to be almost identical in both cases, i.e. the SQUID and the single microbridge. This was explained in our previous work \cite{PhysRevB.84.045103}.

\begin{figure}[tb] 
\center
\includegraphics[width=11cm]{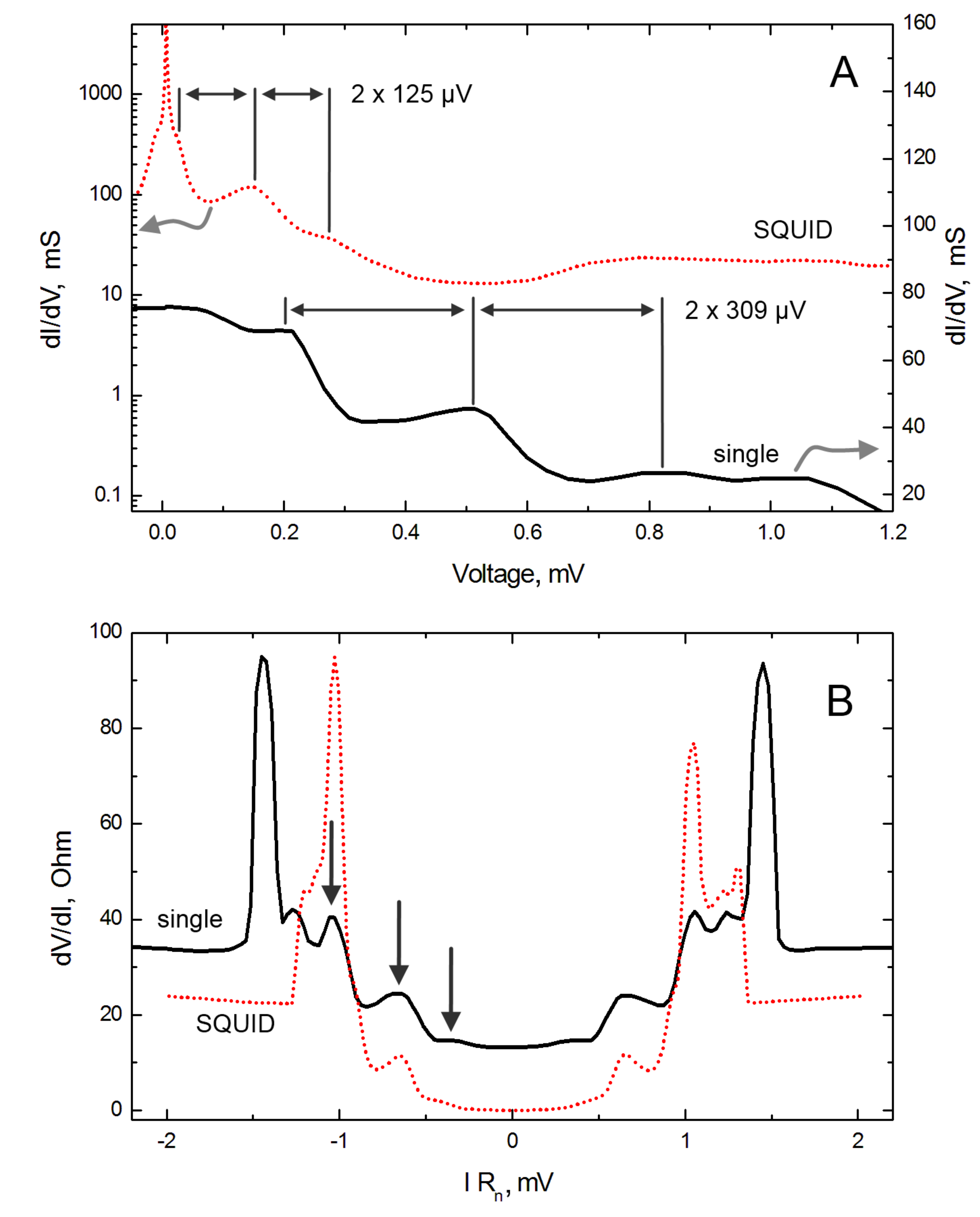} 
\caption{Typical dynamic resistance/conductance characteristics of the single {\cci} microbridge (solid lines) and the SQUID (dotted lines) measured at T=0.3~K at zero external magnetic field. (A) $dI/dV(V)$ representation and (B) the same date in the $dV/dI(IR_n)$ representation, where $R_n$ is the normal state resistance.}
\label{DynamicCharacter}  
\end{figure}

Considering the large values of $L/\xi$ and $W/\xi$ in the {\cci} microbridge, the periodic vortex motion regime should be realizable \cite{RevModPhys.51.101}. As it was shown in the Aslamazov-Larkin (AL) model \cite{AL.1975}, this regime will manifest itself by periodic features in the non-linear conductance curves. In our previous work this behaviour was found for the SQUID's microbridges on the dynamic resistance curves. The $dI/dV(V)$ curves measured in the current driven mode for the SQUID and for the single microbridge are shown in Figure~\ref{DynamicCharacter}(A).
The same date are also shown as $dV/dI(IR_n)$ in Figure~\ref{DynamicCharacter}(B). It should be noted that the $I_c R_n$ is an invariant for Josephson junctions and reflects only essential material properties (see p.200 in Reference~\cite{TinkhamIntro}), hence it is convenient to compare different microbridges in coordinates $dV/dI (I R_n)$. In Figure~\ref{DynamicCharacter}(A) the curve measured for the SQUID demonstrates a sharp Josephson conductance peak around zero voltage bias. When $I>I_c$, a voltage drop along the SQUID's microbridges develops and the first shallow knee appears just above $I_c$, which is linked to the entrance of the first Abricosov vortex pair into the bridge. Three similar features are observed on both the $dI/dV(V)$ and $dV/dI(IR_n)$ curves for the SQUID microbridges, which are associated with three pairs of vortices moving simultaneously within each of the microbridges. As it was mentioned, according to the AL model these features in the dynamic resistance should be periodic with voltage, but not with current or $IR_n$, reflecting the non-linear vortex-vortex interaction. The presented curves for the SQUID follow these AL model predictions, which supports the vortex origin of the conductance features. 

The conductance and resistance curves for the single microbridge demonstrate the same qualitative behaviour as for the SQUID, which suggests an unchanged mechanism of periodic vortex motion within the single microbridge. In contrast to the SQUID data, the single microbridge curve $dI/dV(V)$ in Figure~\ref{DynamicCharacter}(A) demonstrates an only weakly enhanced conductance around zero voltage, which is in both cases the effect of the Josephson current. The currents $I<I_c$, however, are accompanied by a voltage drop, which suggests the presence of dissipative transport channels within the bridge. Up to three periodic features are still recognized on the corresponding plot in Figure~\ref{DynamicCharacter}(A) at $0.2~mV$, $0.51~mV$, and $0.82~mV$ corresponding each to the entrance of a new vortex pair into the bridge. These features are marked by arrows in Figure~\ref{DynamicCharacter}(B). Their slightly non-periodic pattern in the $dV/dI(IR_n)$ plot and periodic pattern in the corresponding $dI/dV(V)$ plot are evident, which is consistent with the AL model. The first two features at small $IR_n$ values coincide with those measured for the SQUID. This suggests a similar vortex dynamics in the single bridge and the SQUID at moderate vortex densities. The third feature is shifted towards larger values of $IR_n$ for the single microbridge when compared with the SQUID. This may be explained by an increased local magnetic penetration depth $\lambda$ enhancing the vortex-vortex repulsion at larger vortex densities. The increase of $\lambda$ may not be surprising, since it is inversely proportional to the density of SC electrons, which may well be decreased due to the mentioned additional dissipation transport channels. At even larger values of $IR_n$ the current density within the bridge reaches its critical value for {\cci} \cite{PhysRevB.70.020506}, above which the dynamic resistance characteristics become essentially flat, which is expected for microbridges.
\section{Conclusion} 
A microbridge was prepared from an individual growth domain of a $c$-axis oriented {\cci} thin film. The microbridge demonstrated DC Josephson effects which was confirmed by several typical characteristics, namely a magnetic field driven periodic voltage modulations $\Delta V(B)$, a temperature dependence of the Josephson critical current $I_c(T)$ following the Ambegaokar-Baratoff model, and by the periodic vortex motion predicted for microbridges of large dimensions when compared to characteristic SC length scales. The observed asymmetry of $\Delta V(B)$ also suggests large dimensions of the microbridge. However, the behaviour of $I_c(T)$ did not demonstrate any of the expected behaviour predicted for microbridges, but, in contrast, followed the behaviour of a simple Josephson junction in the tunneling regime. Josephson effect studies on microbridges of unconventional superconductors can provide a valuable access to the symmetry of the order parameter. Our work indicates that this approach is also applicable to moderate quality thin films of heavy fermion superconductors. 
\section{Acknowledgments} 
This work was supported by the Deutsche Forschungsgemeinschaft (DFG) through grant No. HU 752/3-3 and by the Beilstein-Institut, Frankfurt am Main, Germany, within the research collaboration NanoBiC.
 




\bibliographystyle{model1a-num-names}







\end{document}